\documentclass{article}
\usepackage{spconf,amsmath, graphicx}
\usepackage{comment}
\usepackage{cite}
\usepackage{color}
\usepackage{enumerate}
\usepackage{array}
\usepackage{breqn}
\usepackage{amssymb}
\usepackage{multirow}
\usepackage{arydshln}
\title{Building English ASR model with regional language support}
\name{{Purvi Agrawal $^{\dagger}$, Vikas Joshi $^{\dagger}$, Bharati Patidar, Ankur Gupta, Rupesh Kumar Mehta}
 \thanks{The manuscript was prepared in February, 2023. The author Dr. Purvi Agrawal was full-time employee as Applied Research Scientist at Microsoft at the time of publication.\\
$^\dagger$ denotes equal contribution.}}
\address{Microsoft Corporation \\ \small{Email: purvia@iisc.ac.in, \{vijoshi, bpatidar, angup, rupeshme \}@microsoft.com}}

\ninept
\begin{document}
\ninept
\maketitle

\begin{abstract}
In this paper, we present a novel approach to developing an English Automatic Speech Recognition (ASR) system that can effectively handle Hindi queries, without compromising its performance on English. We propose a novel acoustic model (AM), referred to as SplitHead with Attention (SHA) model, features shared hidden layers across languages and language-specific projection layers combined via a self-attention mechanism. This mechanism estimates the weight for each language based on input data and weighs the corresponding language-specific projection layers accordingly. Additionally, we propose a language modeling approach that interpolates n-gram models from both English and transliterated Hindi text corpora. Our results demonstrate the effectiveness of our approach, with a $69.3\%$ and $5.7\%$ relative reduction in word error rate on Hindi and English test sets respectively when compared to a monolingual English model.
 

\end{abstract}



\section{Introduction}
\label{sec:intro}
We analysed the queries on our Indian English (en-IN) automatic speech recognition (ASR) system and observed the following:
\begin{itemize}
    \item About one-third of the queries were entirely in regional languages. For example, ``Kesariya ka bhakti gaana dikhaiye” is one such query, where both the entity and language of conversation are in Hindi. Note that the Hindi words are represented in Latin characters for wider readability. The English translation of this query is ``show Kesariya's bhakti song".
    \item Most entertainment queries had regional language movie or song names.
    \item More than $30\%$ of the call center queries involved code-switching to regional languages for all utterances.

\end{itemize}
It is becoming increasingly evident that users may interact in regional languages, even if they have selected English as their preferred language. While current voice assistants can recognize entities from different languages, they still face challenges when the entire query, including the language of conversation, is in another language. Some voice assistants now support bilingual queries, allowing users to interact interchangeably in English and Hindi without explicitly setting the language of the conversation. This system is achieved by using two monolingual models in tandem and selecting the output language based on the decision of the language identifier (LID) \cite{joshi2021multiple}. However, using such a system much higher computation cost than monolingual models, as both monolingual models must produce recognition outputs, increasing the cost of recognition for the end user. Additionally, the performance of these systems is sensitive to the accuracy of the LID, as the monolingual models perform well in their respective languages but suffer greatly in other languages. The aim of this work is not to build a truly bilingual model that can recognize both languages equally well. Instead, the goal is to improve the performance of the Indian English (en-IN) ASR model on Hindi (hi-IN) queries, bringing it reasonably close to the performance of the Hindi model on Hindi queries. This is to be achieved without increasing the computational costs or reducing its performance on English queries. This model is referred to as the \textit{enhanced} en-IN ASR model. It is more pragmatic and commercially viable, as it can simply replace the existing monolingual en-IN ASR model without affecting the accuracy or computational cost for existing users.

In this work, we propose a novel approach to build an en-IN ASR model that improves performance on Hindi queries without regressing on English queries. Our approach includes a split-head with attention (SHA) acoustic model (AM), which has shared hidden layers and two language-specific projection layers. The output of the language-specific projection layers are weighed by corresponding weights obtained from an attention model and then added to obtain posterior probabilities over the labels, chenones in this case~\cite{le2019senones}. It is important to note that in our approach, both the outputs from language-specific projection layers must correspond to the same set of chenones in order to be added together. This differs from a Shared Hidden Layer modeling approach~\cite{ SHL_2}, where each language can have its own set of output labels. 
The language model (LM) is obtained by interpolating an en-IN n-gram LM with the one obtained from transliterated Hindi text via domain adaptation approach. Additionally, we propose to use contextual rather than word-based transliteration approach. With these innovations in AM and LM, our enhanced en-IN ASR model shows $69.3\%$ and $5.7\%$ word error rate relative (WERR) reduction over a conventional en-IN model on Hindi and English test sets, respectively.

\section{Relation to prior work}
\label{sec:relevant_works}
In the acoustic modeling , transfer learning, multilingual and multi-dialect modeling are widely used in hybrid ASR systems\cite{MLT_1,swietojanski2012unsupervised, MLT_2,  TL_1, TL_2, Scanzio-MultisoftmaxFirstPaper, SHL_2, Seltzer-MTLPhonemeRecog, EL_2, Das-MultiDialectEnsemble}. Transfer learning methods~\cite{TL_1,TL_2} leverage a well trained AM from high-resource language to bootstrap the target AM. A natural extension is to train multilingual model \cite{Scanzio-MultisoftmaxFirstPaper,SHL_2,Seltzer-MTLPhonemeRecog}, by using data from multiple languages to build robust seed model. However, such a model has inferior performance compared to their monolingual counterparts, as reported in \cite{punjabi2020streaming, van2010pooling}. The approach of multiple monolingual models in tandem and selecting an output based on LID results into a costly system for the end user \cite{gonzalez2014real}. Hence, our goal in this paper is to design an enhanced AM in hybrid framework without regressing on primary language and increase in user cost.

In LM,
similar to work discussed in \cite{DBLP:journals/corr/abs-2103-10730}, a core aspect of our approach is the transliteration of regional language data to the Latin script. However, in order to keep our LMs lightweight, fast and easily modifiable, we do not build neural LMs, and instead, focus on interpolation-based techniques. Such techniques \cite{kneser1993dynamic, klakow1998log, allauzen2011bayesian, vu2012first, raju2018contextual} have been around for a while, but to the best of our knowledge, it has not been used to support a secondary language (Hindi, in this case) while not regressing on the primary language (English, in this case). 
There are some approaches that focus purely on the problem of code-mixing, such as \cite{garg2017dual, garg-etal-2018-code, chandu-etal-2018-language}. Some of these approaches require code-mixed training data or additional information such as language at run-time. It is important to note that our goal is different from these with our goal being to build a good English LM with the additional support of Hindi, not necessarily interspersed with each other or occurring in the same utterance.

The rest of the paper is organized as follows. Section \ref{sec:proposedAM} discusses the proposed AM in detail, followed by proposed LM in Section \ref{sec:proposedLM}. Section \ref{sec:experiments} discusses the experiments and results, followed by Summary in the last Section.

\section{Proposed AM: SplitHead with Attention (SHA)} {\label{sec:proposedAM}}
In this section, we will discuss the model architecture, data preparation and training strategy for our proposed approach. \subsection{Model architecture}
We introduce Splithead with Attention (SHA) model, a novel architecture designed for multilingual and code-mixed speech recognition. It can be broadly divided into two parts:


\textbf{(a) SplitHead model}: The SplitHead AM features language-specific projection/output layers and shared hidden layers, as illustrated in Figure \ref{fig:blockDiag}(a). By having a separate output layer for each language, the model can learn language-specific characteristics. At the same time, shared layers benefit from robust training using a large amount of data. This approach is similar to Shared Hidden Layer (SHL) training but with one key difference. In SHL training, each language can have its respective chenones (the output units or smallest speech units modeled). However, in our proposed approach, we restrict all languages to use the same chenones as the primary language. This enables us to combine posteriors from multiple languages in the next step.

\textbf{(b) SplitHead with attention:} 
The final model must produce a single output vector representing posterior probabilities over all chenones. However, the SplitHead model discussed above generates two outputs (one for each language via its respective projection layer) for each input frame. Therefore, we combine these two outputs from different languages using an attention mechanism to produce a single weighted output vector \cite{patil2023streaming}, as illustrated in Fig. \ref{fig:blockDiag}(b). Let our AM takes a chunk of $L$ frames from $\mathbf{X}=\{\mathbf{x}^t,...,\mathbf{x}^{t+L}\}$ to produce hidden representations $\mathbf{h}^t$ to $\mathbf{h}^{t+L}$ as shown in Fig. \ref{fig:blockDiag}(c), where $L$ is referred as no. of lookahead frames. The self-attention module~\cite{vaswani2017attention} takes these hidden representations $\{\mathbf{h}^t,...,\mathbf{h}^{t+L}\}$ as input to produce language-specific scalar weights $w_{en}^t$ and $w_{hi}^t$. This attention mechanism is part of the same acoustic model and is trained jointly with rest of the AM parameters. Therefore, the output of the proposed model $P(\mathbf{y}^t|\mathbf{X})$ is a weighted combination of posterior probabilities from different languages $\mathbf{h}_{en}^t$ and $\mathbf{h}_{hi}^t$, followed by softmax as, 
\begin{equation}
\mathbf{h}^t = w_{en}^{t}\mathbf{h}_{en}^t + w_{hi}^{t}\mathbf{h}_{hi}^t
\end{equation}
\begin{equation}
P(\mathbf{y}^t|\mathbf{X}) = Softmax(\mathbf{h}^t)
\end{equation}

\begin{center}
    \begin{figure}[t!]
        \centering
        \includegraphics[scale=0.35]{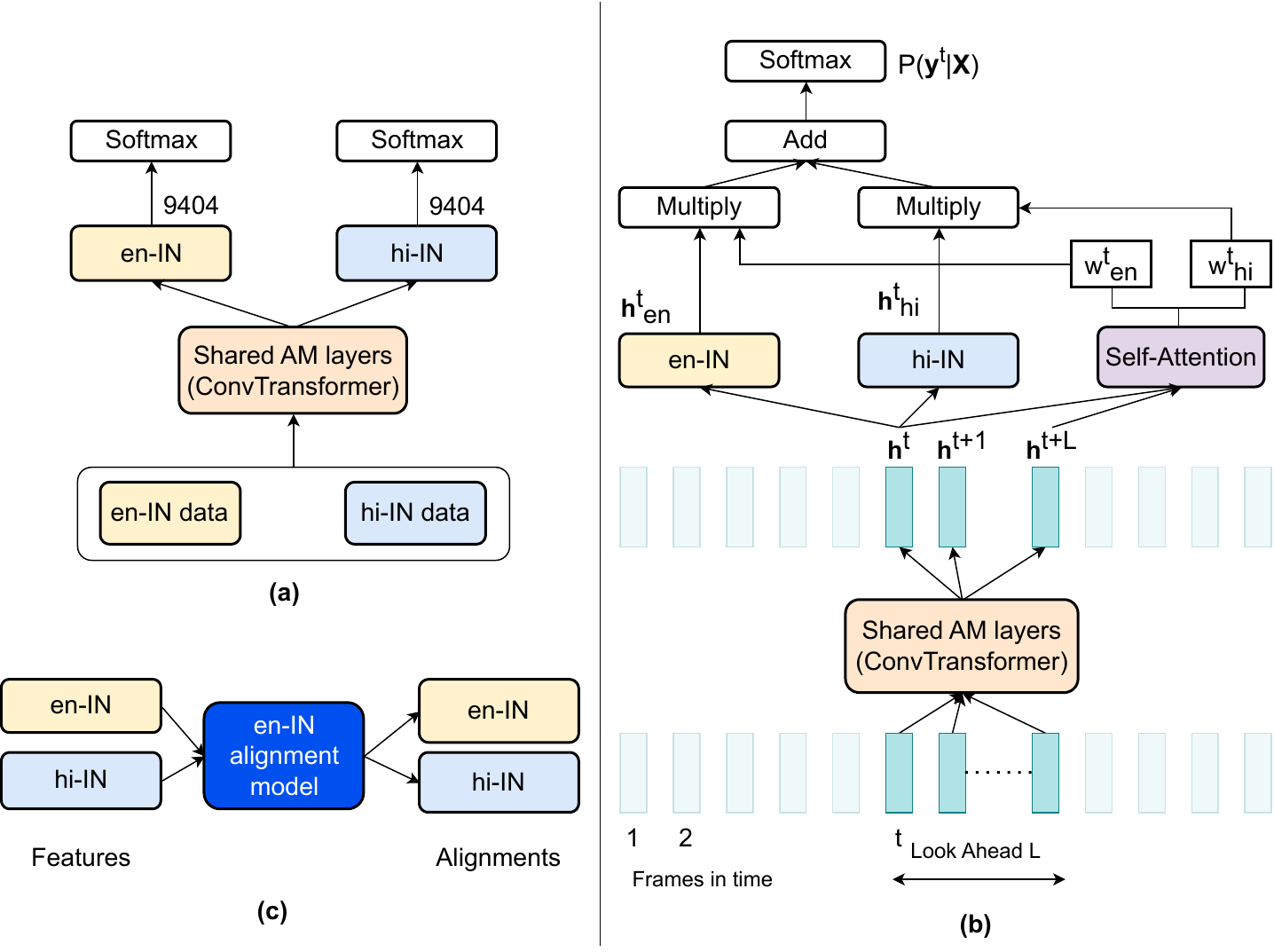}
        \vspace{-0.2cm}
        \caption{Block schematic of (a) SplitHead model approach, (b) Proposed SplitHead with attention AM approach, (c) Data preparation.}
        \label{fig:blockDiag}
    \vspace{-0.2cm}
    \end{figure}
    \vspace{-1cm}
\end{center}

\subsection{Data preparation and training strategy}
The data preparation and training strategy for the SHA model differ from those of a conventional AM and is described in this section. Data preparation involves obtaining alignments using a pre-trained ASR model. Unlike the conventional approach of using language-specific alignment models, we use an en-IN alignment model to align both English and Hindi data, as shown in Figure \ref{fig:blockDiag}(c). This ensures that we have the same chenone label set for both languages, which is a specific requirement of the SHA model. We choose the alignment model of the primary language (en-IN in this case) to align both the languages to maintain parity with the primary language. To align Hindi data with the en-IN alignment model, we transliterate Hindi data into English. Alternatively, we could use phone mapping\cite{madhavaraj2019data}, but this would require linguistic expertise to obtain such a mapping.


The SHA model is trained with cross-entropy (CE) loss in following stages to effectively train all its components:
\begin{enumerate}
    \item \textbf{Train SingleHead model:} We begin by training a single projection layer model using pooled data from both the languages. This model is referred to as the ‘SingleHead’ or Data-Pooled (DP) model.
    \item \textbf{Train SplitHead model:}
    We convert the SingleHead model into a SplitHead model by initializing the projection layers for both languages with the projection layer from the SingleHead model. The shared layers are initialized with the corresponding shared layers from the SingleHead model~\cite{Joshi2020}. The shared hidden layers are trained using data from all languages while the projection layers are trained using only language-specific data. As a result, each output head is observed to perform well on its respective language (discussed in Section~\ref{sec:experiments}).

    \item \textbf{Train attention only:} 
     Starting with a well-trained SplitHead model, we train the SHA model by updating only the self-attention module parameters and freezing all other parameters. The attention model is trained using data from all languages, including code-mixed utterances so that it learns to generate appropriate weights.
    \item \textbf{Full model training:} 
    Finally, we train the entire SHA model including shared layers, all projection layers and the self-attention model using pooled data from all the languages.
\end{enumerate}

\subsubsection{Knowledge distillation from teacher model}
We also explore teacher student training to further improve the performance of the SHA model. Our teacher model is an ensemble of two models: a) A data-pooled SingleHead \textit{non-streaming} model trained with the same en-IN and hi-IN data, and b) The SHA \textit{streaming} student model. The inclusion of SHA student model ensures that it does not deviate significantly from its original state. We empirically choose $0.4$ weight to non-streaming teacher and $0.6$ to streaming SHA model as the best performing criteria. We then finetune the SHA model with teacher-student loss ($L_{TS}$) as weighted combination of cross entropy loss ($L_{CE}$) and Kullback–Leibler divergence (KLD) loss ($L_{KLD}$) as shown below,
\vspace{-0.05cm}
\begin{equation}
L_{TS} = (1-w_{KLD})L_{CE} + w_{KLD}L_{KLD}
\end{equation}
\vspace{-0.05cm}
where $w_{KLD}=0.95$ represents the weight of the KLD loss. 

\section{Proposed LM} 
\begin{center}
    \begin{figure}[t]
        \centering
        \includegraphics[scale=0.53]{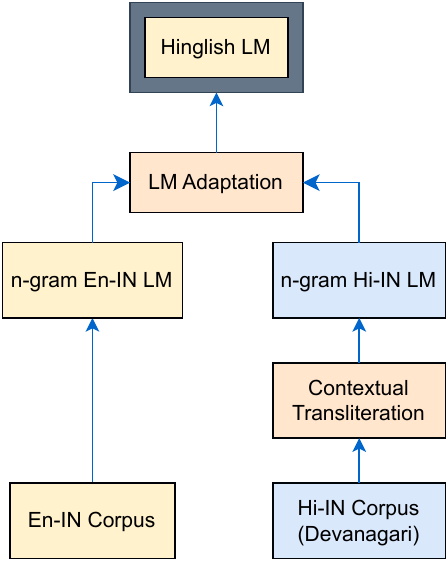}
        \vspace{-0.1cm}
        \caption{Block schematic of building Hinglish LM.}
        \label{fig:LMinASR}
        \vspace{-0.25cm}
    \end{figure}
\end{center}

{\label{sec:proposedLM}}
This section outlines our approach towards building the LM for \textit{enhanced} en-IN ASR, with the goal of creating an English ASR model that can also support Hindi and is referred to as \textit{Hinglish} LM. Our approach employs interpolation, a widely used technique for adapting LMs to specific domains \cite{kneser1993dynamic, klakow1998log, allauzen2011bayesian}, to achieve effective language adaptation.  As a result, the LM can support Hindi without negatively impacting performance on the English test set. Figure \ref{fig:LMinASR} depicts our approach to build the Hinglish LM with the following steps.  

\textbf{Data collection}: The en-IN and hi-IN LM corpus are collected from a wide variety of sources popular in India and are akin to its people and culture. We collect data from public domains such as \textit{news websites}, \textit{Wikipedia}, and \textit{social media} to capture the diverse range of dialects and language patterns in informal and everyday communication. We also use conversational transcriptions from \textit{interviews} and \textit{podcasts} available in the public domain. Entity-specific regional data is mined from public domains, knowledge graphs and Wikipedia. Entities are mined from different categories including but not limited to - \textit{Movies}, \textit{TV shows}, \textit{Songs}, \textit{Celebrities}, \textit{Sports Persons}, \textit{Politicians}, \textit{Companies}, \textit{Businesses}, \textit{Festivals}, \textit{Geographical locations}, and \textit{common peoples' names}.
   
\textbf{Transliterate from Devanagari script to Latin script}:
    The next step is to transliterate hi-IN LM data from Devanagari script into Latin script as shown in Fig.\ref{fig:LMinASR}. We use Microsoft Transliteration service for the same. This ensures that the output of Hindi queries is presented in Latin characters, which is necessary as we are replacing the existing en-IN ASR model.
    Latin script is widely accepted and commonly used format for representing Hindi in everyday scenarios with predominance on social media platforms as well. Use of transliterated Hindi corpus in Latin script, ensures output in Latin script for both English and Hindi queries. We explored these two approaches for transliteration as discussed below:

        \begin{enumerate}
            \item \textbf{Word-based transliteration:} 
        Each word in the hi-IN LM corpus is first transliterated to Latin English and is stored as a look-up table. This is followed by word-by word replacement in the the hi-IN corpus using the look-up table. This approach is cost-effective as only one call is made to the transliteration service per word in the vocabulary. We observed that it improved the baseline En-IN model to recognize Hindi patterns and entities, but it failed to capture the long-range semantic relationship within the text. It also caused inaccuracies and inconsistencies, especially when different words could represent the same sound depending on the context. For eg., English word “pay” and Hindi transliterated word “pe” sound exactly the same, but are used in different contexts.
            \item \textbf{Contextual transliteration: } 
        Here, instead of word, complete sentence  is transliterated by passing entire sentence to the transliteration API. This approach is more expensive than the word-based approach but is observed to preserve the context and meaning of the text. We show in the discussion of results that this approach helped to achieve lower WER than the word based transliteration.
        \end{enumerate}


\textbf{LM Adaptation}: The final step involves building n-gram LMs for both en-IN and transliterated hi-IN corpora and interpolating them as shown in Figure \ref{fig:LMinASR}. We experimented with different interpolation weights and obtained optimal results with en-IN and hi-IN LMs weights as $0.9$ and $0.1$, respectively. This relative scoring ensures that we do not compromise performance in English while improving performance in Hindi.

\vspace{-1cm}
\section{Experiments and Results}
\label{sec:experiments}
\subsection{Data and experimental set-up}
We conduct experiments on English (en-IN), Hindi (hi-IN) and Hindi-English (Hinglish) code-mixed utterances. We use approximately $20.6K$ hours of training data consisting of $8K$ hours of Hindi, $10K$ hours of English and rest of code-mixed data. The test set consists of $13070$ Hindi utterances and $37890$ English utterances (comprising of conversational, phrasal and dictation scenarios). The Hindi test sets are transliterated to Latin as the en-IN model produces Latin output only.

We use $80$-dimensional log mel filterbank features, computed every $10$ milliseconds (ms). These $80$ dimensional features are stacked to form $160$ dimensional features.
A  base architecture of convolutional transformer (CTFM) with $18$ conv-transformer blocks is used to train baseline and the proposed models \cite{gulati2020conformer}. Models are trained with Adam optimizer and with $32$ GPUs. A hybrid DNN-HMM speech recognition system is trained for all experiments with $9404$ chenones, the tied context-dependent graphemes \cite{le2019senones}. 
\begin{center}
    \begin{table}[t] {\label{tab:results_CE}}
        \centering
        \vspace{-0.2cm}
        \caption{Word error rate (\%) for baseline and different stages of the proposed ASR at CE and SE stage, respectively.
        }
        \resizebox{\linewidth}{!}{
         \begin{tabular}{|c|l|c|c|c|}
        \hline
        No. & AM Approaches & LM & en-IN & Trans hi-IN \\ \hline
        \multicolumn{5}{|l|}{CE stage} \\ \hline
        I & en-IN (baseline) & en-IN &  13.71 & 58.63\\ 
        II & en-IN & Hinglish & 13.73 & 29.80 \\
        III & SingleHead & Hinglish & 14.17 & 19.97 \\
        IV & SplitHead & Hinglish & 12.95 & 20.04/26.9 \\
        V & SHA (Pretrain attn only) & Hinglish &  12.82 & 20.48 \\
        VI & SHA (PretrainAttn+fullTrain) & Hinglish &  13.66 & 18.37  \\
        VII & SHA + KLD loss & Hinglish & 12.93 & 17.95 \\
         \hline \hline
        \multicolumn{5}{|l|}{SE stage} \\ \hline
        VII & SHA + KLD loss & Hinglish & 11.33 & 14.63 \\ \hline
        \end{tabular}
        }
        
    \end{table}
\end{center}
\vspace{-1cm}
\subsection{Discussion of Results}
Table $1$ shows the ASR performance in word error rate (WER \%) on en-IN test sets and transliterated hi-IN test sets over proposed approach at different stages. The baseline (I) is a monolingual en-IN ASR model with AM and LM trained on en-IN data only. At CE stage, the Hinglish LM with en-IN AM (II) gives significant improvements with $49.2\%$ WERR reduction over transliterated hi-IN test sets without regressing on the en-IN test set. We next discuss the performance of the AM at different stages. The SingleHead AM trained in data-pooled fashion with en-IN and hi-IN data (III) shows $33\%$ WERR reduction on hi-IN sets compared to II, while it introduced $3.2\%$ regression on en-IN test sets.

The SplitHead AM with two heads (IV) shows $8.6\%$ and $-0.3\%$ WERR reduction over SingleHead model (III) on en-IN and hi-IN test sets, respectively. Here we obtain the outputs from the respective language heads. Note that we evaluated hi-IN test sets over each head [hi/en head: $20.04/26.9$] and observed hi-head having significant gains over en-head as expected. The proposed SHA AM after pretraining + full training (VI), shows $3.6\%$ and $8\%$ improvement over SingleHead model (III) on hi-IN and en-IN test sets respectively. Finally, the SHA model trained with CE+KLD loss (VII) using ensemble teacher model gives best performance with $8.8\%$ and $10.1\%$ improvement over SingleHead model (III) on en-IN and hi-IN test sets, respectively. Compared to the monolingual en-IN baseline (I), the proposed SHA + KLD loss along with Hinglish LM (VII) gives $5.7\%$ and $69.3\%$ WERR reduction on en-IN and transliterated hi-IN test sets, respectively. Results also show the consistent improvements with SHA + KLD loss approach (VII) at sequence discriminative training (SE) stage trained by maximizing mutual information (MMI).
Our future work includes improving the hi-IN WER to come further closer to monolingual hi-IN model without impacting en-IN WER in order to become a truly bilingual model.


\begin{center}
    \begin{figure}[t] {\label{fig:attn_wts_dist}}
        \centering
        \includegraphics[scale=0.57]{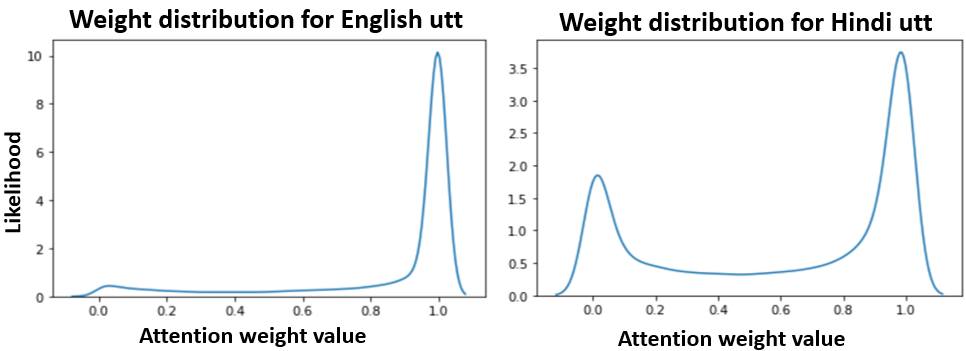}
        \caption{Distribution of self-attention weights for `en' head and `hi' head over English and Hindi utterances, respectively. }
    \end{figure}
\end{center}
\vspace{-1.2cm}
\subsubsection{Analysis of attention weights}
Figure $3$ shows the distribution of attention weights obtained from the self-attention model. For each set of utterances (English and Hindi), the attention weights are estimated over all the frames and we fit a multi-modal Gaussian distribution.
As seen from the Figure, $w_{en}$ peaks near $1.0$ for English utterances and $w_{hi}$ peaks near $1.0$ for Hindi utterances. The  English distribution is more flat compared to Hindi one and the Hindi distribution has an extra small peak near $0.0$. It can be attributed to lot of code-mixed and code-switched utterances with usage of English words in our Hindi data. In summary, the analysis of attention weights show that it conforms to our initial hypothesis of attention model behaving like an implicit language identification system, where $w_{en}$ is close to $1.0$ for English words/frames and $w_{hi}$ is close to $1.0$ for Hindi words/frames.




\subsubsection{Varying number of language-specific transformer layers}
Table $2$ shows the effect of using different number of language-specific transformer blocks \textit{n}. Note that in SplitHead architecture, projection layers (after $18$ shared transformer blocks) are always language-specific. Hence, \textit{n} $=0$ implies no language-specific transformer block, \textit{n} $=1$ implies $17$ shared blocks and $1$ language-specific block for each language, and so on. As can be observed, we see some regressions with \textit{n} $=1$ and \textit{n} $=2$. However, it starts to improve again with \textit{n} $=4$ followed by regressions upon further increase. While we expected more gains with increased language specific layers, we did not further investigate as the size of the model increased with more language specific layers. Hence, we choose \textit{n} $=0$ for all our experiments with optimal trade-off between performance and model size.
\begin{center}
    \begin{table}  {\label{tab:results_diff_langSpecifcLayers}}
        \centering
        \caption{Word error rate (\%) for SplitHead approach with varying no. of language-specific transformer blocks (\textit{n}) over En-IN test sets.
        }
        
        \resizebox{0.78\linewidth}{!}{
        \begin{tabular}{|c|c|c|c|c|c|}
        \hline
        \textit{n} & 0 & 1 & 2 & 4 & 8 \\ \hline
        WER (\%) & 12.95 & 13.40 & 13.21 & 12.93 & 13.1 \\\hline
        
        \end{tabular}
        }
        \vspace{-0.3cm}
    \end{table}
\end{center}
\vspace{-0.9cm}
\subsubsection{Comparison of word-based vs. contextual transliteration}
Here we report the gains from Hinglish LMs prepared using the two different transliteration approaches discussed earlier in Section~\ref{sec:proposedLM}.
We observed $32.89\%$ WERR reduction from word-based LM approach and a significant WERR reduction of $49.2\%$ from contextual transliteration, over monolingual English baseline. Hence, we build the Hinglish LM using contextual transliteration for our models. 

\vspace{-0.1cm}
\subsubsection{No. of parameters and Latency}
The proposed SHA model has $5.6\%$ increase in parameters compared to baseline monolingual model. We conducted latency tests on CPU machines with identical configuration and did not observe any noticeable change in the latency. This is attributed to the negligible change in the amount of  additional computation due to extra projection layer and an attention module, compared to latency of forward pass, viterbi decoding and lookahead frames. 
\vspace{-0.1cm}
\section{Summary} 
This work proposes a new approach to develop an enhanced English ASR with added Hindi support without compromising its English performance. To the best of our knowledge, several methods have been proposed to build bilingual and multilingual models, but this is the first methodological approach to improve the existing ASR model’s performance on a secondary regional language. We find this approach to be a more programmatic solution considering the mix of end users with some requiring high English model accuracy while others require English with reasonable accuracy on the regional language. Towards this end, we proposed a novel SHA model with shared hidden layers and language-specific projection layers combined via a self-attention mechanism. We also proposed training the LM by interpolating n-gram LMs from an English text corpus and a contextually transliterated Hindi text corpus. The proposed approach shows a $69.3\%$ WERR reduction over the monolingual English baseline on the secondary language Hindi.

\bibliographystyle{IEEEbib}
\bibliography{refs}

\end{document}